\documentclass[11pt]{cernrep}
\usepackage{graphicx}
\usepackage{here}
\usepackage{amsmath}
\newcommand{\Vek}[1]{\mbox{\boldmath${#1}$}}

\begin{document}
\title{A NEW METHOD FOR THE HIGH-PRECISION ALIGNMENT OF TRACK DETECTORS}
\author{Volker Blobel and Claus Kleinwort}
\institute{Institut f\"ur Experimentalphysik,
Universit\"at Hamburg,  and Deutsches Elektronensynchrotron DESY, Hamburg, Germany}
\maketitle
\begin{abstract}
Track detectors in high energy physics experiments require an
accurate determination of a large number of alignment parameters.
A general method has been developed, which allows the determination
of up to several thousand alignment parameters in a simultaneous
linear least squares fit of an arbitrary number of tracks.
The sensitivity of the method is demonstrated in an example of
the simultaneous alignment of a 56-plane drift chamber
and a 2-plane silicon tracker. About 1400 alignment parameters
are determined in a fit of about fifty thousand tracks.
\end{abstract}

\section{ALIGNMENT PROBLEMS OF TRACK DETECTORS} 

\begin{figure} \begin{center}
\includegraphics*[bb = 88 272 524 546, clip=true, width=12cm]{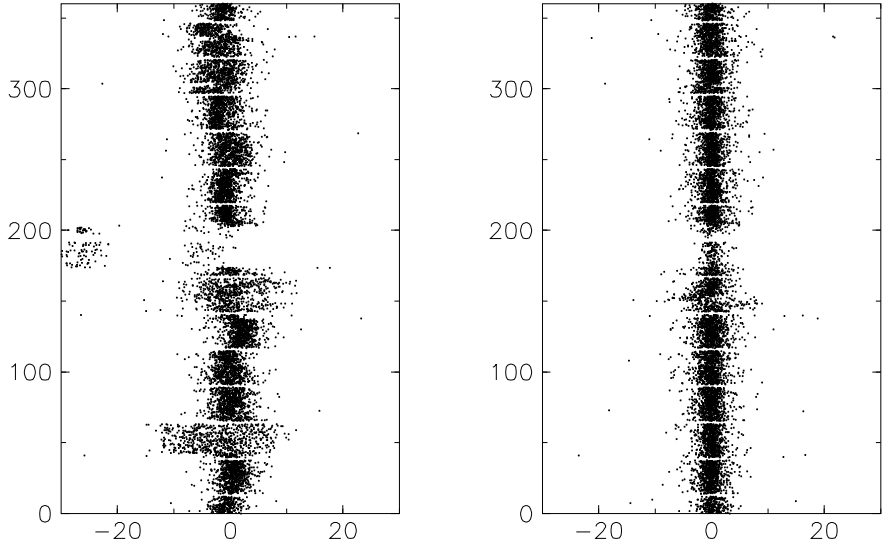} 
 \begin{minipage}[b]{12cm}
\caption{Residuals between tracks and hits of a silicon tracker are
shown as a function of the azimuthal angle (vertical axis, in degrees)
before and after an alignment. The residual on the horizontal axis is
given in units of 10 $\mu$m.  \label{fig:corra}}
\end{minipage}
\end{center}
\end{figure}

Alignment problems for large detectors, especially track detectors, 
in particle physics often require the
determination of a large number of alignment parameters, typically of the
order of $100$ to $1000$. Alignment parameters for example define the
accurate space coordinates and orientation of detector components.
In addition drift chambers  require the determination of the 
drift velocity $v_{\text{drift}}$ and of local variations of the 
drift velocity, and of the Lorentz angle. Figure \ref{fig:corra}
shows an example of residuals in a silicon vertex track detector
before and after an alignment. 

In the alignment usually special alignment measurements are combined with
data of particle reaction, typically tracks from physics interactions and
from cosmics. In this paper the alignment parameters are called \emph{global}
parameters; they contribute to all the data. Usually these parameters 
are \emph{corrections} to default values and are thus relatively small,
and the value zero is the initial approximation.
Parameter sets valid for a
single track are called \emph{local} parameters, examples are track slopes and
curvatures.    
  
An approximate alignment method is to perform least squares fits on the data
\emph{ignoring} initially the alignment parameters. The \emph{deviations}
between the fitted and measured data, the \emph{residuals},
are then used to determine the alignment
parameters afterwards. Single fits depend only on the
small number of local parameters
like slope or curvature of the track, and are easy to perform.
This approximate method however is not a statistically correct method,
because the 
(local) fits depend on a wrong model (namely: ignoring the global
parameters), and the result are biased. In addition the results may
be unstable due to detector inefficiencies.
The adjustment of parameters
based on the observed (biased) residuals will then result in
biased alignment parameters. If these alignment parameters are 
applied as corrections in repeated fits, the remaining
residuals will be reduced, as desired. However, the fitted parameters
will still be biased, even more, if this approximate procedure is applied
iteratively.  The simple residual method is also unable to determine
correlated parameters, for example a geometrical shift of a detector
position and a shift in drift velocity in a drift chamber.  

A better method is an overall fit, with all the (global)
alignment parameters and local parameters, perhaps from thousands or
millions of events, determined simultaneously.
The most important general method for the determination of 
parameters from measured data is the linear least squares model. 
It is usually stable, fast and accurate, requires no initial values
of parameters and is
able to take into account all the correlations between different 
parameters, even if there are many parameters.  
There is of course a practical limit in the number of parameters in
simultaneous fits, because of space and time limitations. 
In an alignment fit
the size of the vectors and matrices in the least
squares solution will be large, of the order of $10^4$ or $10^6$. 
With standard methods such a solution is impossible.  

In this paper a special method of solution for this kind of
problems is derived. 
Due to the special structure of the matrices with one set of 
global parameters and many sets of local
parameters the problem can in fact be reduced to a solvable size,
without making any approximations. The rather general method is realized in a
program called {\bf Millepede} \cite{millepede}.

\section{STANDARD LEAST SQUARES}

\noindent {\bf Linear models.}
In a linear model the measured quantity $z_k$ has an expectation value
which is a \emph{linear} combination of the parameters $\Vek{a}$ with fixed 
factors (e.g. from geometry) which are combined to a vectore $\Vek{d}_k$.
The 
difference between the measured value and the expectation value is
the residuum $r_k$: 
\begin{equation}
        z_k = \Vek{a}^T \cdot \Vek{d}_k + r_k \; .
\end{equation}
The measured data are assumed to be \emph{independent}; the result of 
a measurement $z_k$ does not depend on any other measurement.
Thus the covariance matrix of the measured data is a diagonal matrix, which
simplifies the treatment. The accuracy of the measurement $z_k$ is 
given by the variance $\sigma^2_k$ or the standard deviation 
$\sigma_k$. It is assumed, that the precision is at least approximately
known. 

The solution of the method of least squares 
is determined by the 
minimum of the  weighted  sum of the square of the residuals
\begin{equation}
 S(\Vek{a})  = \sum_k \frac{\left(z_k -
                \Vek{a}^T \cdot \Vek{d}_k 
                    \right)^2}{\sigma_k^2}  \; , 
\end{equation} 
where the sum (index $k$) is over all measurements. This sum
 is a quadratic function of the parameters $\Vek{a}$, and the 
solution is obtained from a set of linear equations (so-called normal 
equations of least squares), which can be written in matrix form   
\begin{equation}  \label{eq:solemio}
             \Vek{C} \Vek{a} = \Vek{b}
\quad \quad \text{with} \quad 
\Vek{C}  = \sum_k w_k \left( \Vek{d}_k \cdot  \Vek{d}_k^T \right)
\quad \quad \text{and} \quad
\Vek{b}   = \sum_k w_k   z_k  \Vek{d}_k 
\end{equation}
with a symmetric $n$-by-$n$ matrix $\Vek{C}$ and a right-hand-side
$n$-vector $\Vek{b}$; the  elements of $\Vek{C}$ and $\Vek{b}$ are
 sums with \emph{independent} contributions from each measurement
(because the measured data are independent).
There is no limitation on the number of measured data; it is not necessary
to keep all input data in memory, because the complete
information from a single
measurement is contained in the contributions to $\Vek{C}$ and $\Vek{b}$,
which have a fixed size. The solution vector is 
$\Vek{a} = \Vek{C}^{-1} \Vek{b}$
with the inverse of the matrix $\Vek{C}$, which, by error propagation, is
equal to the covariance matrix of the parameters $\Vek{a}$. 

What are the limits for the number $n$ of parameters, if the matrix equation
is solved by standard methods? 
The symmetry property of the matrix $\Vek{C}$  may be used
to store only the $(n^2+n)/2$ elements on and above the diagonal, 
and also to reduce
the computer time to almost half the time  for a non-symmetric matrix.
Matrix inversion is an explicit process and requires a CPU time
proportional to $n^3$ for a $n$-by-$n$  matrix. 
For a $1000$-by-$1000$ symmetric matrix in double precision
(about 4 Mbyte) the CPU time is a few minutes
on a standard PC. The time ($\propto n^3$)  and space ($\propto n^2$)
requirements imply at present  a practical
limit  of the size of the matrix (and number of parameters) between
$n = $ 1~000 and $n = $ 10~000.

The accuracy of the computation of the inverse matrix is a critical point, 
and often computation in double precision is required, especially for
large matrices. The accuracy 
is generally low for a large condition number of the matrix (the condition 
number is defined as the ratio of the largest to the smallest
eigenvalue of the matrix). From the interpretation of the inverse 
matrix as covariance matrix it is clear, that the condition number will be
large if the correlation between parameters is high. The matrix is 
singular (one or more eigenvalues are zero) in case of a complete 
correlation between parameters. In this case and also in cases of 
missing data (e.g. dead channels in an alignment fit) a standard matrix
inversion program will fail.  
A singular value decomposition (SVD) would 
allow to recognize the cases of zero or small eigenvalues, which 
would \emph{destroy} the full solution. 
The standard method in case of zero or very small eigenvalues is to 
set  their inverse to zero; this means essentially to remove parameters
or linear combinations of parameters from the solution (i.e.  
the resulting parameter values are zero); if the parameters are in fact
\emph{corrections} this is clearly acceptable. 
In the program {\bf Millepede} \cite{millepede}  
a similar, but simpler and faster strategy
is used to avoid the effects of a bad condition number. This strategy
also works in cases, where due to missing input data\footnote{Large 
detectors may have dead channels, which means missing data.}   
one or more parameters are undefined; the parameters will become
zero in this case. The strategy allows in addition to remove certain 
parameters from the least squares fit without the necessity to 
reorganize the numbering. 

\noindent{\bf Solution by partitioning.} 
The special structure of the matrix $\Vek{C}$ to be inverted may allow
to move the limit of solvable problems further (see also \cite{blolohr}).
 Below the matrix
 $\Vek{C}$ is partitioned into submatrices, and the vectors 
$\Vek{a}$ and $\Vek{b}$  are partitioned into two subvectors;  then the 
matrix equation becomes
\begin{equation}
   \left(
       \begin{array}{ccc|c}
 &           & &         \\
 &    \Vek{C}_{11} & & \Vek{C}_{12}  \\
 &           & &         \\   \hline
 &     \Vek{C}_{21} & & \Vek{C}_{22}
       \end{array}
      \right)     
   \left( \begin{array}{c}
  ~\\   \Vek{a}_1  \\  ~\\  \hline \Vek{a}_2
          \end{array} \right)  =
   \left( \begin{array}{c}
   ~\\   \Vek{b}_1  \\  ~\\  \hline  \Vek{b}_2
          \end{array} \right)  \; ,
\end{equation}
where the submatrix $\Vek{C}_{11}$ is a
  $p$-by-$p$ square
matrix and the submatrix $\Vek{C}_{22}$ is a $q$-by-$q$
square matrix, with $p+q=n$. 
If the sub-vector $\Vek{a}_1$ would not exist, the solution for the 
sub-vector $\Vek{a}_2$ would be defined by the matrix equation
\begin{equation}
                \Vek{C}_{22} \; \Vek{a}_2^{*} = \Vek{b}_2
\quad \quad \quad \text{with solution} \quad
           \Vek{a}_2^{*} =   \Vek{C}_{22}^{-1} \;  \Vek{b}_2 \; ,
\end{equation}
where the star indicates the special character of this solution.
Only the $q$-by-$q$ 
sub-matrix $\Vek{C}_{22}$ has to be inverted. 
The solution $\Vek{a}_2^{*}$ of course differs from the final solution
$\Vek{a}_2$ and is called the \emph{local} solution.

Now, having inverted the sub-matrix $\Vek{C}_{22}$, the submatrix 
of the complete inverse matrix $\Vek{C}$ correponding to the upper left
part $\Vek{C}_{11}$, called $\Vek{B}$, is  obtained by the formula
\begin{equation}   \label{eq:modinv}
  \Vek{B} =  \left( \Vek{C}_{11} - \Vek{C}_{12} \Vek{C}_{22}^{-1}
     \Vek{C}_{12}^T\right)^{-1}      \; ,
\end{equation}
which requires in addition to some matrix products the 
calculation of the inverse of a $p$-by-$p$ sub-matrix.  
With this matrix $\Vek{B}$ the solution of the whole matrix equation can 
written in the form
\begin{equation}
   \left( \begin{array}{c}
  ~\\   \Vek{a}_1  \\  ~\\  \hline \Vek{a}_2
          \end{array} \right) =
   \left(
       \begin{array}{ccc|c}
 &           & &         \\
 &    \Vek{B} &  & - \Vek{B} \Vek{C}_{12} \Vek{C}_{22}^{-1}  \\
 &           & &         \\   \hline
 &   - \Vek{C}_{22}^{-1} \Vek{C}_{12}^T \Vek{B} 
 & & \Vek{C}_{22}^{-1} - \Vek{C}_{22}^{-1}
 \Vek{C}_{12}^T \Vek{B} \Vek{C}_{12} \Vek{C}_{22}^{-1} 
       \end{array}
      \right)     
   \left( \begin{array}{c}
   ~\\   \Vek{b}_1  \\  ~\\  \hline  \Vek{b}_2
          \end{array} \right) \; ,
\end{equation}
which can be used to obtain the complete result ($\Vek{a}_1$ and
$\Vek{a}_2$). Note that to solve the matrix equation two smaller matrices,
a symmetric $p$-by-$p$  and a symmetric $q$-by-$q$ matrix,  have to be
inverted. 
The sub-vector $\Vek{a}_1$ is obtained by the product 
\begin{equation}  \label{eq:finala1} 
       \Vek{a}_1 = \Vek{B} \Vek{b}_1 -
          \Vek{B} \Vek{C}_{12}
     \Vek{C}_{22}^{-1} \Vek{b}_2   
=    \Vek{B}  \left( \Vek{b}_1 - \Vek{C}_{12} \Vek{a}_2^* \right) 
\; , 
\end{equation}
which is simplified by using the special local solution
 $\Vek{a}_2^* = \Vek{C}_{22}^{-1} \Vek{b}_2 $.
If the interest is the determination of this sub-vector $\Vek{a}_1$
only, while
the sub-vector $\Vek{a}_2$ is not needed, then only the equation
\eqref{eq:finala1}
has to be solved. Some computer time can be saved by this 
method, especially if the matrix $\Vek{C}_{22}^{-1}$ is easily calculated 
or already known before. In special cases the method can be applied
repeatedly, 
and therefore problems with a huge number of parameters become
solvable. 
The method is not an approximation, but is exact and it takes into account
all the correlations.
This type of application is discussed in more detail in the next chapter.

It seems that this possibility of removing unnecessary parameters,
but still getting the correct solution for the remaining parameters
in least squares problems was already known and has
been used in the nineteenth century; one example is \cite{schreiber}.

\noindent{\bf Constraints.}
If there are explicit relations between the parameters of a least squares
fit, then they should be explicitly taken into account in the solution.
The case of a single linear constraint between parameters of the form
$\Vek{f}^T \cdot \Vek{a} = \Vek{f}_0$ is assumed here.
The Lagrange multiplier method is the standard method to include constraints
in a least squares solution; one additional vector $\Vek{\lambda}$ is
introduced, with one element per constraint.
The equation \eqref{eq:solemio} is modified by the additional parameters
to the form 
\begin{equation}
   \left(
       \begin{array}{ccc|c}
 &           & &         \\
 &    \Vek{C} & & \Vek{f}  \\
 &           & &         \\   \hline
 &     \Vek{f}^T & & \Vek{0}
       \end{array}
      \right)
   \left( \begin{array}{c}
  ~\\   \Vek{a}  \\  ~\\  \hline \Vek{\lambda}
          \end{array} \right)  =
   \left( \begin{array}{c}
   ~\\   \Vek{b}  \\  ~\\  \hline  \Vek{f}_0 
          \end{array} \right)  \; ,
\end{equation}
where the matrix on the left hand side is still symmetric.
\emph{Linear} least squares problems with \emph{linear}
constraints can be solved direcly, without iterations and without
the need for initial values of the parameters.

\section{LEAST SQUARES WITH GLOBAL AND LOCAL PARAMETERS}

\noindent{\bf Local parameters.}
 Local parameters are denoted by 
$\alpha_j, \; j=1, \, 2,\ldots \nu$. A set of measurements $z$ is 
considered. A single value $z$ from the set is ideally given by a linear 
expression $z = \alpha_1 \cdot \delta_1 +
             \alpha_2 \cdot \delta_2 +
             \ldots 
             \alpha_{\nu} \cdot \delta_{\nu}$,
depending on the parameters $\alpha_j$ and on (known) constant factors
 $\delta_j$.
More complete, the dependence can be written as 
\begin{equation}  \label{eq:zetka2}
         z_k = \alpha_1 \cdot \delta_{1k} +
             \alpha_2 \cdot \delta_{2k} +
             \ldots 
             \alpha_{\nu} \cdot \delta_{\nu k} 
           = \sum_{j=1}^{\nu} \alpha_j \cdot \delta_{jk}   \; ,
\end{equation}
with an additional index $k$ indicating the $k$-th measured value. 

As an example the  measurement of  a straight particle track is
considered, where each $z_k$ is a coordinate measurement at plane $k$.
The local parameters are intercept ($\alpha_1$) and slope ($\alpha_2$)
of the particle track. If the distance of the plane is denoted by $S_k$,
the linear track model can be written as
\begin{displaymath}
      z_k = \alpha_1 \cdot 1 + \alpha_2 \cdot S_k   \; .
\end{displaymath}
From a least squares fit of the model to the set of measured $z$-values
the local parameters intercept and slope can be determined.
The formalism of least squares  allows to take into account
the different precision of the different measurements, by assigning a weight 
 to each measurement of  $w_k = 1/\sigma_k^2$.  
The least squares solution, minimizing the weighted sum of squared
residuals, is defined by the so-called normal equations, which can be 
written in terms of matrices
\begin{equation}   \label{eq:solvenormal}
        \Vek{\Gamma} \, \Vek{\alpha} = \Vek{\beta}\, ,
 \quad \quad \text{solved by} \quad
        \Vek{\alpha} = \Vek{\Gamma}^{-1} \Vek{\beta}   \; . 
\end{equation}   
 The elements of the matrix
$\Vek{\Gamma}$ and the vector  $\Vek{\beta}$ are formed by sums
\begin{equation} \label{eq:normaleq}
     \Gamma_{ij} = \sum_k w_k \cdot \delta_{ik} \delta_{jk}
\quad \quad \quad \quad \quad \quad
     \beta_j = \sum_k w_k \cdot z_k \delta_{jk}   \; .
\end{equation}
These formulas are valid in the case of \emph{uncorrelated} $z$-measurements
(uncorrelated measurements are assumed thoughout this paper).
The inverse matrix $\Vek{\Gamma}^{-1}$ is the covariance matrix of the 
parameter vector  $\Vek{\alpha}$.

\noindent{\bf Global parameters.}
Now global parameters are considered, which contribute to \emph{all} the 
measurements. They are assumed to be \emph{alignment parameters} here, 
 although the formalism is general and can also be applied to other problems.  
Aligment parameters   
 are defined to represent
\emph{corrections} to ideal (design) values; this parametrization adds
further terms to the equations \eqref{eq:zetka2}
\begin{equation}   \label{eq:zetall}
         z =
\underbrace{a_1 \cdot d_1 +
             a_2 \cdot d_2 +
             \ldots 
             a_n \cdot d_n}_{\text{global parameters}} 
           + 
\underbrace{\alpha_1 \cdot \delta_1 +
             \alpha_2 \cdot \delta_2 +
             \ldots 
             \alpha_{\nu} \cdot \delta_{\nu}}_{\text{local parameters}} 
           = \sum_{i=1}^n a_j \cdot d_j   + 
            \sum_{j=1}^{\nu} \alpha_j \cdot \delta_j   \; .
\end{equation}        
Usually only few terms containing global parameters are nonzero for 
a single $z$ measurement. 
In the following it is assumed that, as written in equation 
\eqref{eq:zetall}, the dependence on the global parameters is
\emph{linear}. A non-linear relationship for the global parameters
would require iterations, starting with reasonable initial values 
for the parameters, and assuming a linearized expression 
in each iteration. 

\noindent{\bf The simultaneous fit of global and local parameters.}
In the following it is assumed that there is a set of $N$ partial measurements.
Each partial measurement, with index $i$, depends on local parameters
$\Vek{\alpha}_i$, and all of them depend on the global parameters.   
In a simultaneous fit of all global parameters plus local parameters
from $N$ subsets of the data there are in
total $(n+N\cdot\nu)$ parameters, and the standard solution requires the 
solution of $(n+N\cdot\nu)$ equations with a computation proportional to 
$(n+N \cdot \nu)^3$. Below it is shown, that the 
problem can be reduced to $n$ equations with a computation proportional
$n^3$.   

Generalizing the formalism of equations \eqref{eq:zetka2}
to the complete measurement, starting from equations of the type
of \eqref{eq:zetall}, one obtains a system of normal equations with large 
dimensions, as is shown in equation \eqref{eq:huge}.   
The matrix on the left side of equation \eqref{eq:huge} 
 has, from each partial measurement, three types of contributions. 
The first part is a contribution of a symmetric matrix $\Vek{C}_i$, of 
dimension $n$ (number of global parameters). All the matrices  $\Vek{C}_i$
are added up in the upper left corner of the big matrix of the normal
equations. The second contribution is the symmetric matrix 
$\Vek{\Gamma}_i$ (compare equation \eqref{eq:normaleq}),
 which gives a contribution to the diagonal of the  big matrix
 and is depending only on the $i$-th partial measurement.  
The third contribution is a rectangular matrix $\Vek{G}_i$, with 
a row number of $n$ (global) and a column number of $\nu$ (local). 

There are two contributions to the vector of the normal equations, 
$\Vek{b}_i$ for the global and $\Vek{\beta}_i$ for the local parameters.
The complete matrix equation is given by
\begin{equation}  \label{eq:huge}  \renewcommand{\arraystretch}{1.2}
  \left(
       \begin{array}{ccc||ccc|c|ccc}
 &           & &  & & & & & &         \\
 &   \sum \Vek{C}_i  & &  & \cdots & &  \Vek{G}_i & & \cdots &  \\
 &           & &  & & & & & &        \\   \hline \hline
 &   & &  &  & & & & & \\
 &  \vdots & & & \ddots & & 0 & & 0 & \\
 &   & & & &  & & & & \\  \hline
   &    \Vek{G}^T_i &  &  & 0 & & \Vek{\Gamma}_i & &0 &  \\ \hline
 &   & & & & & &  & & \\
 &  \vdots & & & 0 & & 0 & & \ddots & \\
 &   & & & & & & & &  \\ 
       \end{array}
      \right)    
. \left( \begin{array}{c}
  \\  \Vek{a} \\  \\  \hline  \hline 
  \\ \vdots \\  \\  \hline
  \Vek{\alpha}_i \\ \hline
  \\ \vdots \\  \\ 
         \end{array} \right)
= 
. \left( \begin{array}{c}
  \\  \sum \Vek{b}_i \\  \\  \hline \hline
  \\ \vdots \\  \\  \hline
  \Vek{\beta}_i \\ \hline
  \\ \vdots \\  \\ 
         \end{array} \right)  
\end{equation}
In this matrix equation the matrices  $\Vek{C}_i$,
$\Vek{\Gamma}_i$, $\Vek{G}_i$
and the vectors  $\Vek{b}_i$ and $\Vek{\beta}_i$
contain contributions from the $i$-th partial measurement.
Ignoring the global parameters one could solve the normal equations
 $ \Vek{\Gamma}_i \Vek{\alpha}_i^* = \Vek{\beta}_i$
for each partial measurement separately by 
\begin{equation}  \label{eq:ignore}
 \Vek{\alpha}_i^* = \Vek{\Gamma}_i^{-1} \Vek{\beta}_i \, .
\end{equation}
The complete system of normal equations has a special structure, with many
vanishing sub-matrices. The only connection between the local parameters of
different partial measurements is given by the sub-matrices
 $\Vek{G}_i$ und $\Vek{C}_i$,

The aim of the fit is solely to determine the global parameters; 
final best parameters of the local parameters are not needed.
The matrix of equation \eqref{eq:huge} is written in a partitioned form.
The general solution can also be written in partitioned form.
Many of the sub-matrices of the huge matrix in equation \eqref{eq:huge} 
 are zero and this has the effect, that the 
formulas for the sub-matrices of the inverse matrix are very simple.  
By this procedure the normal equations 
\begin{equation}   \label{eq:nsb}   \renewcommand{\arraystretch}{1.2}
  \left(
       \begin{array}{ccc}
 &           &      \\
 &    \Vek{C'}  &  \\
 &           &      \\ 
       \end{array}
      \right)    
 \left( \begin{array}{c}
  \\  \Vek{a} \\  \\    
         \end{array} \right)
= 
  \left( \begin{array}{c}
  \\  \Vek{b'}  \\  \\  
         \end{array} \right)   \; ,
\end{equation}
are obtained, which only contain the global parameters, with a 
modified matrix $\Vek{C'}$ and a vector $\Vek{b'}$,
\begin{equation}  \label{eq:nsc}
  \Vek{C'} =  \sum_i \Vek{C}_i - \sum_i
 \Vek{G}_i \Vek{\Gamma}_i^{-1} \Vek{G}_i^T 
 \quad \quad 
  \Vek{b'} =    \sum_i \Vek{b}_i - \sum_i
 \Vek{G}_i \left( \Vek{\Gamma}_i^{-1} \Vek{\beta}_i\right)    \; .
\end{equation}
This set of normal equations contains explicitly only the global
parameters; implicitly it contains, through the correction matrices, 
the complete information from the local parameters, influencing the
fit of the global parameters. The parentheses in equation \eqref{eq:nsc}
represents the solution for the local parameters, ignoring the global 
parameters.  The solution
  $\Vek{a} = \Vek{C'}^{-1}\,  \Vek{b'}$ 
represents the solution vector $\Vek{a}$ with covariance matrix 
  $\Vek{C'}^{-1}$. 
The solution is direct, no iterations are required. 
Iterations may be necessary for other reasons, namely 
\begin{itemize}
\item the equations depend \emph{non-linearly} on the global
parameters; the equations have to be linearized; 
\item  the data contain outliers which have to be removed in a sequence
of cuts, becoming narrower during the iteration;
\item the accuracy of the data is not known before, and has to be 
determined from the data (after the alignment).
\end{itemize}

\noindent{\bf Application.}
A very important point in problems discussed above is the 
definition of the set of global parameters. 
It is essential to define them in a way which is sensitive to the 
data used to determine them. Computing the inverse of a huge (symmetric)
matrix can be a delicate numerical problem. The matrix will be 
singular or close to a singular matrix, if one or more of the 
global parameters are not really determined (and constrained) by the 
data, and this could destroy the whole determination. This will happen 
in case of a strong correlation between different global parameters.

It is wise to start with a few global parameters and to observe the
correlations between them, before adding more global parameters.  For example
if too many global parameters are used in a detector alignment, the whole
detector may \emph{shift} and \emph{rotate} freely in the fit.
In those cases one could either fix certain global parameters or one has
to introduce constraints (e.g. average displacement and rotation equal
to zero). 

Experience with the method shows that it is essential to make 
a simultaneous alignment of all track detector components, which are also
used for the measurement of tracks, while independent
internal alignment of the     
single track detector components may not lead to a good overall result. 
It is also important to use, if possible, different data sets 
simultaneously. For example in a track detector alignment
high-momentum cosmic muons may be useful, which traverse the whole 
detector and thus relate (and constrain) different parts of the detector. 
But this should be done simultaneously with tracks from physics events,
because these events finally require a good alignment. 

The method described here has been realized in the program 
{\bf Millepede} \cite{millepede}, written in Fortran77.
It provides a set of subroutines for the mathematical methods and 
allows to adapt the method to a particular problem.
The program allows to adjust the dimension of
vectors and matrices via \verb/PARAMETER/-statements 
and to introduce linear constraints.
Throughout the program use is made of the sparse character of the vectors
and matrices in order to reduce the execution time. The program 
includes a method for outlier rejection, which in practice may be 
essential in alignment problems especially if some detector 
components have a bad initial alignment. This feature requires 
to iterate, realized with an intermediate data file, which is 
written during the first iteration and read during the other iterations.  

Sometimes the model underlying the alignment is \emph{nonlinear} 
and also constraints may be nonlinear. The standard method to treat 
these problems is linearization: the nonlinear equation is replaced
by a linear equation for the correction of a parameter (Taylor 
expansion); this requires a good approximate value of the parameter.

\section{ALIGNMENT OF THE CENTRAL TRACK DETECTORS IN THE H1 EXPERIMENT}

The example described is the alignment in the $r \varphi$-plane 
perpendicular to the beam line of a 56-plane drift chamber and
a 2-plane silicon vertex detector in the H1 detector at HERA \cite{h1}.
Both are cylindrical detectors; the drift chamber has a length ($z$) of 
about 2 m, and extending from 20.3 cm to 84.4 cm in radius $r$. 
The silicon vertex detector \cite{cst}
has two planes around the beampipe made of 0.15 mm Aluminium and 0.9 mm
carbon fiber. The drift chamber and the silicon vertex detector have a
 $r \varphi$ resolution of about 150 $\mu$m and 15 $\mu$m, respectively. 
These detectors are
interspersed by additional chambers for the $z$-measurement.
 
\begin{table}
 {\footnotesize
\begin{tabular}{|r|r|lrl|}
\hline 
row. & number &  parameter & $\sigma$  & unit \\  \hline
1    &   2 &  $\Delta x$             & 1  & $\mu$m   \\  
2    &   2 &  $\Delta x/\Delta z_r$  & 2  & $\mu$m   \\
3    &   2 &  $\Delta y$             & 1  & $\mu$m   \\ 
4    &   2 &  $\Delta y/\Delta z_r$  & 2  & $\mu$m   \\
5    &   2 &  $\Delta \varphi$        & 10 & $\mu$rad \\     
6    &   2 &  $\Delta \varphi/\Delta z_r$   & 10 & $\mu$rad \\
7    &   2 &  $\Delta \alpha_{\text{Lor}}$   &100 & $\mu$rad \\
8    &   2 &  $\Delta v_{\text{drift}-}/v_{\text{drift}}$    & $10^{-5}$ &  \\ 
9    &   2 &  $\Delta v_{\text{drift}+}/v_{\text{drift}}$    & $10^{-5}$ &  \\ 
10   &   2 &  $\Delta T_0 \times v_{\text{drift}}$  & $< 1$ & $\mu$m \\ \hline
11   &   2 &  wire staggering in wire plane & few  & $\mu$m \\
12   &   2 &  wire staggering perp wire plane & few & $\mu$m \\
13   &   2 &  sagging in wire plane & few  & $\mu$m \\
14   &   2 &  sagging perp. wire plane & few & $\mu$m \\
15   & 180 &  $\Delta v_{\text{drift}}/v_{\text{drift}}$ per cell half 
                & few &  $10^{-4}$ \\
16   & 112 &  $\Delta v_{\text{drift}}/v_{\text{drift}}$ per layer half
                & few  & $10^{-4}$  \\  
17   & 330 &  $\Delta T_0\times v_{\text{drift}} $ per group & 10  &  $\mu$m  \\
18   &  56 &  wire position in driftdir. per layer & 10  & $\mu$m \\
19   &  56 &  $\Delta T_0 \times v_{\text{drift}}$ per layer & 10 & $\mu$m \\
20   &  56 &  $\Delta x_W$ per layer & few 10 & $\mu$m  \\ 
21   &  112 & $\Delta v_{\text{drift}}/v_{\text{drift}}$ for $I_e/50$ mA &
              few $10^{-4}$ & \\
              \hline 
22   &  90 &  $\Delta v_{\text{drift}}/v_{\text{drift}}$ per layer &
              few $10^{-4}$ &  \\
23   &  90 &  $\Delta y_W$ per layer & few 10 & $\mu$m  \\ 
24  &  90 &  $(\Delta y_W)^2$ per layer$^2$ & few 10 & $\mu$m  \\ \hline \hline
25   &  64 &  $\Delta$ in ladder & few  & $\mu$m  \\
26   &  64 &  $\Delta$ perp. ladder & few  & $\mu$m  \\
27   &  64 &  rel. $\Delta$ in ladder ($z_r$) & few  & $\mu$m  \\
28   &  64 &  rel. $\Delta$ perp. ladder ($z_r$) & 10 & $\mu$m  \\
29   &  64 &  rel. $\Delta$ perp. ladder ($\varphi$) & few  & $\mu$m  \\ \hline
\hline  \end{tabular}}
\hfill
\begin{minipage}[b]{6cm}%
\caption{Alignment parameters determined in the {\bf Millepede} fit.
\label{aliparm}
The {\bf Millepede} accuracy is given by one standard deviation $\sigma$. 
The different parameters are discussed in the text.}\end{minipage}
 \end{table}

The central jet chambers CJC1 and CJC2 (\cite{h1})
with an active length of 2200 mm and an outer radius of 844 mm
 have
in total 2640 anode sense wires 
parallel to the beam with two adjacent cathode planes (wires)
shaping the drift field. A jet chamber cell extends azimuthally from the 
sense wire plane to both adjacent cathode wire planes, and radially extends 
over the full radial span of CJC1 and CJC2 with no further subdivision. 
The jet cells are tilted by about $30^{\circ}$ such that in the presence
of the magnetic field (1.1 Tesla) the ionization electrons  drift
approximately perpendicular to high momentum vertex tracks. 

The silicon vertex detector CST consists of two cylindrical layers
of double sided, double metal silicon sensors read out by a custom designed
analog pipeline chip. The two layers of the CST are formed from 12 and 20 
faces at radii of 5.75 cm and 9.75 cm, respectively. One face or 'ladder'
consists of six silicon sensors and aluminium nitride hybrids at each end.
A double layer of carbon fiber strips with a total thickness of 700 
$\mu$m and a height of 4.4 mm is glued to the edges. The position of the
ladders in a layer are shifted tangentially to ensure an overlap in 
$r \varphi$ of adjacent active areas. The active length in $z$ is 35.6 cm   
for both layers. 

Data taken with the H1 detector are reconstructed online and the 
reconstruction modules also determine important parameters like 
average drift velocities, Lorentz-angle and $T_0$ of the CJC1 and CJC2. 
In addition beam parameters like the $r \varphi$ position of the
primary interaction point are recorded.  

In the alignment procedure, using the program {\bf Millepede}, a large 
number of parameters is determined; these parameter are in general
corrections to detector parameters and represent
\emph{small} corrections.
It turned out that a \emph{common} alignment of the drift chamber and the 
silicon detector is essential, after an internal detector alignment,
which already determines reasonable starting values for alignment parameters. 
For each of the 
drift chambers CJC1 and CJC2  14 global parameters representing
an overall shift or tilt are introduced. Local variations
of the drift velocity $v_{\text{drift}}$ for cells halfs and layers halfs
are observed, which are parametrized by 180 + 112 corrections,
which change with the HV configuration.
For each wire group (8 wires) corrections to $T_0$ are introduced
(330 corrections).
For the silicon vertex detector CST an internal alignment has already
been done, using the same techniques, with in total 384 local 
parameters. 
 In the common alignment fit 5 parameters per half ladder
representing shifts are introduced (320 parameters). 
Table \ref{aliparm} contains all parameters.
A far as possible the parameters are defined with the dimension
\emph{length}. For example the time-zero parameters are multiplied
by the mean drift time. Angles are parametrized as shifts relative 
to a normalized length parameter. For example $z_r$ is defined
as $z/z_{max}$ and has a range --1 to +1. The shift $\Delta x$ with 
$z$ is parametrized as $\Delta x/z_r$ and has the dimension length. 

\begin{figure} \begin{center}
\includegraphics[bb= 12 46 532 787, clip=,width=10cm]{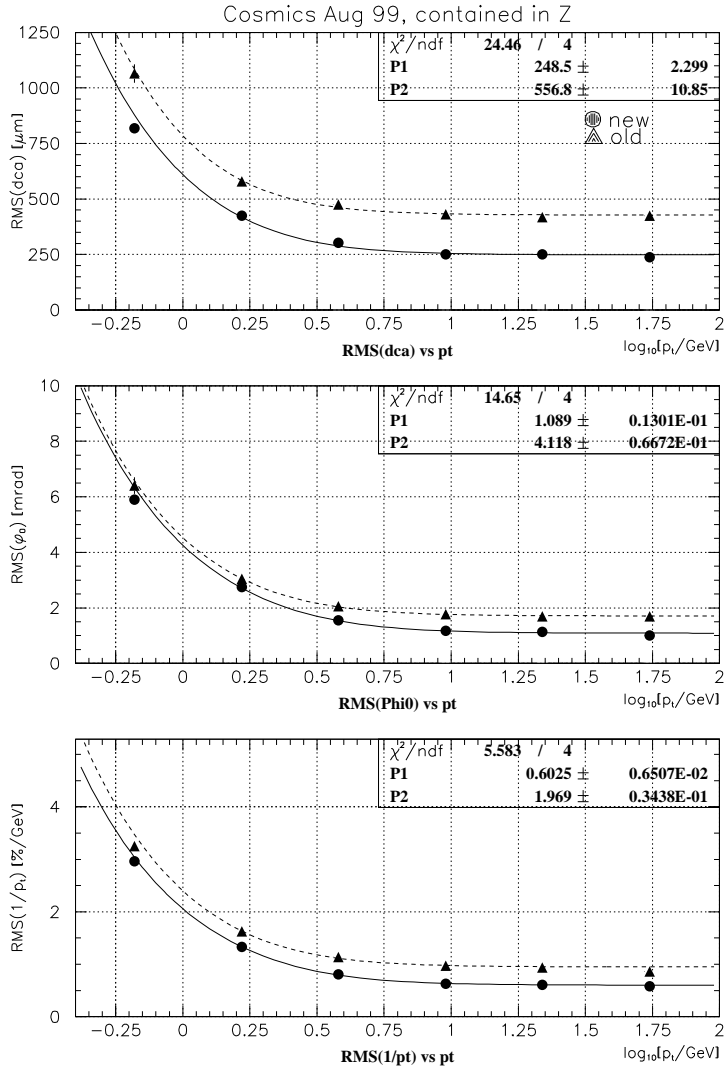}%
\hspace{0.5cm}\begin{minipage}[b]{5.5cm}
\caption{The standard deviations for the three track parameters 
$d_{ca}$ (distance of closest aproach to the beam axis), the 
azimuthal angle $\varphi$ and of the inverse transverse momentum
$1/p_t$ are shown as a function of $\log_{10} p_t/\text{GeV}$,
before (old) and after (new) the improved alignment. 
The standard deviations are determined from the difference 
between the two parts of cosmic tracks. \label{fig:mille2}} 
\end{minipage}
\end{center}
\end{figure}

In addition to physics events (ep interactions) also cosmic 
tracks with and without $B$-field were simultaneously used as 
input to the alignment. With combined fits to both halfs of
 cosmic tracks parts of the detectors opposite 
in $\varphi$ angle are strongly correlated. This allows 
an accurate interpolation from the drift chamber to the 
silicon tracker with an accuracy $\sigma = 30$ to 40 $\mu$m, which 
is of the same order of magnitude as the silicon tracker accuracy.
 In total about 50~000 events 
are used as input. Constraints require e.g. a zero overall shift of the 
detector. The CJC2 is fixed (first five parameters) vs. the cryostat
on the basis of external survey data as reference.  

An overview about the corrections 
determined in the alignment fit is given in Table \ref{aliparm},
which has 29 rows of different corrections. In detail these
corrections are as follows. 
\begin{description}
\item[Rows 1 -- 6] These rows contain geometrical parameters; in
an external survey the positions are measured with an accuracy of 
a few 100 $\mu$m and angles with an accuracy of a few 10 $\mu$rad.  
This accuracy is increased by a large factor in the alignment fit.
\item[Row 6] This parameter 
 is determined from $B=0$ cosmics
(straight line track model).
\item[Rows 11 -- 14] The parameters are determined only as a check. 
Their actual values are determined from hit triplets as a function
of angle $\beta$ and $z_r$ ($\beta$ is the angle between the track
and the normal to the drift direction). 
\item[Row 15,16] The parameters change with the HV configuration.
\item[Row 18,20] Corrections for the wire position in and
 perpendicular to the
drift direction. These corrections are determined in a cross check
with $B=0$ cosmic tracks, otherwise the data from a survey of flanges
are used ($ < 10 \mu$m).  
\item[Row 21] These correction are used to describe the changes of 
drift velocity $v_{\text{drift}}$ by variations of the electron beam
current  $I_e$ due to space charge effects; they are determined once.    
\item[Row 22 -- 24] These parameters change with repair/breakdown of HV
cards.
They are introduced in order to correct for bad cathode HV.
\item[Row 25 -- 29] Corrections for the silicon vertex detector, five per 
half ladder, determined in and perpendicular to ladder, on average 
and as function of $z$ and of $\varphi$.  
\end{description}  
The alignment is done for run ranges and repeated after e.g. a change 
of the HV configuration. Run-to-run variations of parameters
(e.g. the average drift velocity $v_{\text{drift}}$) are taken 
into account by the data from the online reconstruction as mentioned
above. 

\begin{table}[h] \begin{center}
\begin{tabular}{|l|c|c|l|}  \hline
track fit                      
     & $\sigma(d_{\text{ca}}) \; [\text{cm}]$
     & $\sigma(1/p_t) \; [\text{(GeV/c)}^{-1}]$   & remark  \\  \hline  
only 1 track with Si hits            & 0.0209 & $5.58 \cdot 10^{-3}$ &
   drift chamber resolution \\
both tracks with at least 1 Si hit    & 0.0032 & $2.26 \cdot 10^{-3}$ &
   Si tracker resolution \\
both tracks with at least 2 Si hits  & 0.0028 & $2.17 \cdot 10^{-3}$ &
  Si tracker resolution   \\ \hline 
\end{tabular}
\caption{Standard deviations in the parameters  $d_{ca}$ and $1/p_t$,
determined from the
difference between the two parts of cosmic tracks. 
\label{tab:sigmas} } \end{center} \end{table}


The alignment procedure allows to reach the intrinsic resolution 
not only locally but also for complete tracks.
 For the drift chambers CJC1 and CJC2  
the local hit resolution can be determined from local hit triplets 
as a function of drift distance; the minimum is about 120 $\mu$m 
for 1 cm drift distance. The minimum value from track residuals is about 
125  $\mu$m.  For the CST the intrinsic point resolution of 15 $\mu$m from 
hits overlaps is also reached globally. 
Figure \ref{fig:mille2} shows the improvement for track parameters,
determined from data in the comparison of parameters 
between the two parts of cosmic tracks. A significant improvement
is visible for larger track momenta. 
In Table \ref{tab:sigmas} the standard deviations 
 in the parameters $d_{ca}$ and $1/p_t$, determined from the    
difference between the two parts of cosmic tracks, are given
for three different track classes.

\section*{ACKNOWLEDGEMENTS}

We would like to thank the organizers of the conference on
Advanced Statistical Techniques in Particle Physics
for their hospitality and the stimulating atmosphere in Durham.

\end{document}